\documentstyle{article}

\font\msam=msam10 scaled 900

\font\msbm=msbm10 scaled 900
\font\msbm=msbm10
\font\msbmsmall=msbm10 scaled 800
\font\msbmtiny=msbm10 scaled 700
\newfam\msbmfam
\textfont\msbmfam=\msbm
\scriptfont\msbmfam=\msbmsmall
\scriptscriptfont\msbmfam=\msbmtiny
\def\Bbb{\fam\msbmfam\msbm}

\def\varkappa{\mbox{\msbm\char'173}}      

\newtheorem{prop}{Proposition}[section]
\newtheorem{lemma}[prop]{Lemma}

\def\proof{\par{\it Proof}. \ignorespaces}
\def\endproof{{\msam \char'3}\par\smallskip}

\begin{document}

\centerline{\Large On the Asymptotic Expansion of the Solutions}
\centerline{\Large of the Separated Nonlinear Schr\"odinger Equation}

\bigskip
\centerline{\large A.~A.~Kapaev,}
\centerline{\it St Petersburg Department of Steklov Mathematical Institute,}
\centerline{\it Fontanka 27, St Petersburg 191011, Russia,}

\centerline{\large V.~E.~Korepin,}
\centerline{\it C.N.Yang Institute for Theoretical Physics,}
\centerline{\it State University of New York at Stony Brook,}
\centerline{\it Stony Brook, NY 11794-3840, USA}

\begin{abstract}
Nonlinear Schr\"odinger equation with the Schwarzian initial data is 
important in nonlinear optics, Bose condensation and in the theory of 
strongly correlated electrons. The asymptotic solutions in the region
$x/t={\cal O}(1)$, $t\to\infty$, can be represented as a double series in
$t^{-1}$ and $\ln t$. Our current purpose is the description of the
asymptotics of the coefficients of the series.
\end{abstract}

MSC 35A20, 35C20, 35G20


\bigskip
\section{Introduction}

A coupled nonlinear dispersive partial differential equation in $(1+1)$ 
dimension for the functions $g_+$ and $g_-$,
\begin{eqnarray}\label{NLS}
&&-i\partial_tg_+=\frac{1}{2}\partial_x^2g_++4g_+^2g_-,\nonumber
\\
&&i\partial_tg_-=\frac{1}{2}\partial_x^2g_-+4g_-^2g_+,
\end{eqnarray}
called the separated Nonlinear Schr\"odinger equation (sNLS), contains the 
conventional NLS equation in both the focusing and defocusing forms as 
$g_+=\bar g_-$ or $g_+=-\bar g_-$, respectively. For certain physical
applications, e.g.\ in nonlinear optics, Bose condensation, theory of 
strongly correlated electrons, see 
\cite{GK} -- \cite{
DZ}, the detailed information on the long time asymptotics of solutions 
with initial conditions rapidly decaying as $x\to\pm\infty$ is quite 
useful for qualitative explanation of the experimental phenomena.

Our interest to the long time asymptotics for the sNLS equation is inspired
by its application to the Hubbard model for one-dimensional gas of strongly
correlated electrons. The model explains a remarkable effect of charge and 
spin separation, discovered experimentally by C.~Kim, Z.-X.M.~Shen, 
N.~Motoyama, H.~Eisaki, S.~Ushida, T.~Tohyama and S.~Maekawa \cite{KSMEUTM}. 
Theoretical justification of the charge and spin separation include the study 
of temperature dependent correlation functions in the Hubbard model. In the 
papers \cite{GK}--\cite{GIKP}, it was proven that time and temperature 
dependent correlations in Hubbard model can be described by the sNLS equation 
(\ref{NLS}).

For the systems completely integrable in the sense of the Lax representation 
\cite{GGKM, L}, the necessary asymptotic information can be extracted from 
the Riemann-Hilbert problem analysis \cite{ZS1}. Often, the fact of 
integrability implies the existence of a long time expansion of the generic 
solution in a formal  series, the successive terms of which 
satisfy some recurrence relation, and the leading order coefficients can be 
expressed in terms of the spectral data for the associated linear system. For 
equation (\ref{NLS}), the Lax pair was discovered in \cite{ZS2}, while the 
formulation of the Riemann-Hilbert problem can be found in \cite{FT}. As 
$t\to\infty$ for $x/t$ bounded, system (\ref{NLS}) admits the formal solution 
given by 
\begin{eqnarray}\label{as_expansion}
&&g_+=e^{i\frac{x^2}{2t}-(\frac{1}{2}+i\nu)\ln4t}\Bigl(
u_0+\sum_{n=1}^{\infty}\sum_{k=0}^{2n}\frac{(\ln4t)^k}{t^n}u_{nk}\Bigr),
\nonumber
\\
&&g_-=e^{-i\frac{x^2}{2t}-(\frac{1}{2}-i\nu)\ln4t}\Bigl(
v_0+\sum_{n=1}^{\infty}\sum_{k=0}^{2n}\frac{(\ln4t)^k}{t^n}v_{nk}\Bigr),
\end{eqnarray}
where the quantities $\nu$, $u_0$, $v_0$, $u_{nk}$ and $v_{nk}$ are some
functions of $\lambda_0=-x/2t$.

For the NLS equation ($g_+=\pm\bar g_-$), the asymptotic expansion was 
suggested by M.~Ablowitz and H.~Segur \cite{AS}. For the defocusing NLS 
($g_+=-\bar g_-$), the existence of the asymptotic series 
(\ref{as_expansion}) is proven by P.~Deift and X.~Zhou \cite{DZ} using the 
Riemann-Hilbert problem analysis, and there is no principal obstacle to
extend their approach for the case of the separated NLS equation. Thus
we refer to (\ref{as_expansion}) as the Ablowitz-Segur-Deift-Zhou expansion. 
Expressions for the leading coefficients for the asymptotic expansion of the 
conventional NLS equation in terms of the spectral data were found by 
S.~Manakov, V.~Zakharov, H.~Segur and M.~Ablowitz, see 
\cite{M}--\cite{
SA}. The general sNLS case was studied by A.~Its, A.~Izergin, V.~Korepin and
G.~Varzugin \cite{IIKV}, who have expressed the leading order coefficients 
$u_0$, $v_0$ and $\nu=-u_0v_0$ in (\ref{as_expansion}) in terms of the 
spectral data.

The generic solution of the focusing NLS equation contains solitons and 
radiation. The interaction of the single soliton with the radiation was 
described by Segur \cite{S}. It can be shown that, for the generic 
Schwarzian initial data and generic bounded ratio $x/t$, $|c-\frac{x}{2t}|<M$,
$c=const$, where $M$ is small enough, the expansion (\ref{as_expansion}) for
sNLS equation (\ref{NLS}) represents the contribution of the continuous 
spectrum when possible solitons go away from the moving frame. Thus the 
soliton contribution in the considered asymptotic area is exponentially small 
for large $t$ and can be neglected in compare with contributions of the
power-log terms.

Below, we pay our special attention to the coefficients $u_{n,2n}$, 
$v_{n,2n}$, which determine the leading behavior of $g_+$ and $g_-$ as 
$t\to\infty$ in the order $t^{-n}$. For these coefficients as well as for 
$u_{n,2n-1}$, $v_{n,2n-1}$, we find simple {\it exact\/} formulae 
\begin{equation}\label{explicits}
u_{n,2n}=u_0\frac{i^n(\nu')^{2n}}{8^nn!},\quad
v_{n,2n}=v_0\frac{(-i)^n(\nu')^{2n}}{8^nn!},
\end{equation}
and (\ref{un2n-1}) below. We describe coefficients at other powers of $\ln t$
using the generating functions which can be reduced to a system of 
polynomials satisfying the recursion relations, see (\ref{pjqj}),
(\ref{ajbj}). As a by-product, we modify the Ablowitz-Segur-Deift-Zhou 
expansion (\ref{as_expansion}),
\begin{eqnarray}\label{as_modified}
&&g_+=\exp\Bigl[i\frac{x^2}{2t}-(\frac{1}{2}+i\nu)\ln4t+
i\frac{(\nu')^2\ln^24t}{8t}\Bigr]
\sum_{n=0}^{\infty}
\sum_{k=0}^{2n-[\frac{n+1}{2}]}
\frac{(\ln4t)^k}{t^n}\tilde u_{n,k},\nonumber
\\
&&g_-=\exp\Bigl[-i\frac{x^2}{2t}-(\frac{1}{2}-i\nu)\ln4t-
i\frac{(\nu')^2\ln^24t}{8t}\Bigr]
\sum_{n=0}^{\infty}\sum_{k=0}^{2n-[\frac{n+1}{2}]}
\frac{(\ln4t)^k}{t^n}\tilde v_{n,k}.
\end{eqnarray}

\section{Recurrence relations and generating functions}

Substituting (\ref{as_expansion}) into (\ref{NLS}), and equating coefficients 
of $t^{-1}$, we find
\begin{equation}\label{n=0}
\nu=-u_0v_0.
\end{equation}
In the order $t^{-n}$, $n\geq2$, equating coefficients of $\ln^j4t$,
$0\leq j\leq2n$, we obtain the recursion
\begin{eqnarray}\label{ujrec}
&&-i(j+1)u_{n,j+1}+inu_{n,j}=\nu u_{n,j}-
\frac{i\nu''}{8}u_{n-1,j-1}-
\frac{(\nu')^2}{8}u_{n-1,j-2}-\nonumber
\\
&&\hskip-8pt
-\frac{i\nu'}{4}u_{n-1,j-1}'+
\frac{1}{8}u_{n-1,j}''+
\sum_{l,k,m=0\atop l+k+m=n}^n
\sum_{
{\alpha=0,\dots,2l\atop
{\beta=0,\dots,2k\atop
{\gamma=0,\dots,2m\atop
\alpha+\beta+\gamma=j}}}}u_{l,\alpha}u_{k,\beta}v_{m,\gamma},
\\\label{vjrec}
&&i(j+1)v_{n,j+1}-inv_{n,j}=\nu v_{n,j}+
\frac{i\nu''}{8}v_{n-1,j-1}-
\frac{(\nu')^2}{8}v_{n-1,j-2}+\nonumber
\\
&&\hskip-8pt
+\frac{i\nu'}{4}v_{n-1,j-1}'+
\frac{1}{8}v_{n-1,j}''+
\sum_{l,k,m=0\atop l+k+m=n}^n
\sum_{
{\alpha=0,\dots,2l\atop
{\beta=0,\dots,2k\atop
{\gamma=0,\dots,2m\atop
\alpha+\beta+\gamma=j}}}}u_{l,\alpha}v_{k,\beta}v_{m,\gamma},
\end{eqnarray}
where the prime means differentiation with respect to $\lambda_0=-x/(2t)$.

{\it Master generating functions} $F(z,\zeta)$, $G(z,\zeta)$ for the
coefficients $u_{n,k}$, $v_{n,k}$ are defined by the formal series
\begin{equation}\label{FG}
F(z,\zeta)=\sum_{n,k}u_{n,k}z^n\zeta^k,\quad
G(z,\zeta)=\sum_{n,k}v_{n,k}z^n\zeta^k,
\end{equation}
where the coefficients $u_{n,k}$, $v_{n,k}$ vanish for $n<0$, $k<0$ and
$k>2n$. It is straightforward to check that the master generating functions
satisfy the nonstationary separated Nonlinear Schr\"odinger equation in
$(1+2)$ dimensions,
\begin{eqnarray}\label{FG_NLS}
&&-iF_{\zeta}+izF_z=
\bigl(\nu-\frac{i\nu''}{8}z\zeta-\frac{(\nu')^2}{8}z\zeta^2\bigr)F-
\frac{i\nu'}{4}z\zeta F'+\frac{1}{8}zF''+F^2G,\nonumber
\\
&&iG_{\zeta}-izG_z=
\bigl(\nu+\frac{i\nu''}{8}z\zeta-\frac{(\nu')^2}{8}z\zeta^2\bigr)G+
\frac{i\nu'}{4}z\zeta G'+\frac{1}{8}zG''+FG^2.
\end{eqnarray}
We also consider the sectional generating functions $f_j(z)$, $g_j(z)$,
$j\geq0$,
\begin{equation}\label{fjgjdef}
f_j(z)=\sum_{n=0}^{\infty}u_{n,2n-j}z^n,\quad
g_j(z)=\sum_{n=0}^{\infty}v_{n,2n-j}z^n.
\end{equation}
Note, $f_j(z)\equiv g_j(z)\equiv0$ for $j<0$ because $u_{n,k}=v_{n,k}=0$ for
$k>2n$. The master generating functions $F$, $G$ and the sectional generating
functions $f_j$, $g_j$ are related by the equations
\begin{equation}\label{FG_from_fg}
F(z\zeta^{-2},\zeta)=\sum_{j=0}^{\infty}\zeta^{-j}f_j(z),\quad
G(z\zeta^{-2},\zeta)=\sum_{j=0}^{\infty}\zeta^{-j}g_j(z).
\end{equation}
Using (\ref{FG_from_fg}) in (\ref{FG_NLS}) and equating coefficients of
$\zeta^{-j}$, we obtain the differential system for the sectional generating
functions $f_j(z)$, $g_j(z)$,
\begin{eqnarray}\label{fjgjsys}
&&\hskip-20pt
-2iz\partial_zf_{j-1}+i(j-1)f_{j-1}+iz\partial_zf_j=\nonumber
\\
&&\hskip-20pt
=\nu f_j-z\frac{i\nu''}{8}f_{j-1}-
z\frac{(\nu')^2}{8}f_j-z\frac{i\nu'}{4}f_{j-1}'+
z\frac{1}{8}f_{j-2}''+
\sum_{k,l,m=0\atop k+l+m=j}^j f_kf_lg_m,\nonumber
\\
&&\hskip-20pt
2iz\partial_zg_{j-1}-i(j-1)g_{j-1}-iz\partial_zg_j=
\\
&&\hskip-20pt
=\nu g_j+z\frac{i\nu''}{8}g_{j-1}-
z\frac{(\nu')^2}{8}g_j+z\frac{i\nu'}{4}g_{j-1}'+
z\frac{1}{8}g_{j-2}''+
\sum_{k,l,m=0\atop k+l+m=j}^j f_kg_lg_m.\nonumber
\end{eqnarray}
Thus, the generating functions $f_0(z)$, $g_0(z)$ for $u_{n,2n}$, $v_{n,2n}$
solve the system
\begin{equation}\label{f0g0sys}
iz\partial_zf_0=\nu f_0-z\frac{(\nu')^2}{8}f_0+f_0^2g_0,\quad
-iz\partial_zg_0=\nu g_0-z\frac{(\nu')^2}{8}g_0+f_0g_0^2.
\end{equation}
The system implies that the product $f_0(z)g_0(z)\equiv const$. Since
$f_0(0)=u_0$ and $g_0(0)=v_0$, we obtain the identity
\begin{equation}\label{f0g0=-nu}
f_0g_0(z)=-\nu.
\end{equation}
Using (\ref{f0g0=-nu}) in (\ref{f0g0sys}), we easily find
\begin{eqnarray}\label{f0g0}
&&f_0(z)=u_0e^{i\frac{(\nu')^2}{8}z}=
u_0\sum_{n=0}^{\infty}\frac{i^n(\nu')^{2n}}{8^nn!}z^n,\nonumber
\\
&&g_0(z)=v_0e^{-i\frac{(\nu')^2}{8}z}=
v_0\sum_{n=0}^{\infty}\frac{(-i)^n(\nu')^{2n}}{8^nn!}z^n,
\end{eqnarray}
which yield the explicit expressions (\ref{explicits}) for the coefficients 
$u_{n,2n}$, $v_{n,2n}$.

Generating functions $f_1(z)$, $g_1(z)$ for $u_{n,2n-1}$, $v_{n,2n-1}$,
satisfy the differential system
\begin{eqnarray}\label{f1g1sys}
&&\hskip-20pt
-2iz\partial_zf_0+iz\partial_zf_1=
\nu f_1-z\frac{i\nu''}{8}f_0-z\frac{(\nu')^2}{8}f_1-
z\frac{i\nu'}{4}f_0'+2f_1f_0g_0+f_0^2g_1,\nonumber
\\
&&\hskip-20pt
2iz\partial_zg_0-iz\partial_zg_1=
\nu g_1+z\frac{i\nu''}{8}g_0-z\frac{(\nu')^2}{8}g_1+
z\frac{i\nu'}{4}g_0'+f_1g_0^2+2f_0g_0g_1.
\end{eqnarray}
We will show that the differential system (\ref{f1g1sys}) for $f_1(z)$ and
$g_1(z)$ is solvable in terms of elementary functions. First, let us
introduce the auxiliary functions
$$
p_1(z)=\frac{f_1(z)}{f_0(z)},\quad
q_1(z)=\frac{g_1(z)}{g_0(z)}.$$
These functions satisfy the non-homogeneous system of linear ODEs
\begin{eqnarray}\label{p1q1sys}
&&\partial_zp_1=\frac{i\nu}{z}(p_1+q_1)+
\frac{i(\nu')^2}{4}-\frac{\nu''}{8}-
\frac{\nu'}{4}\frac{f_0'}{f_0},\nonumber
\\
&&\partial_zq_1=-\frac{i\nu}{z}(p_1+q_1)-
\frac{i(\nu')^2}{4}-\frac{\nu''}{8}-
\frac{\nu'}{4}\frac{g_0'}{g_0},
\end{eqnarray}
so that
\begin{equation}\label{p1q1+p1sys}
\partial_z(q_1+p_1)=-\frac{(\nu^2)''}{8\nu},
\end{equation}
where we have used the relations (\ref{f0g0=-nu}) and (\ref{f0g0}). Taking
into account the initial condition $p_1(0)=q_1(0)=0$, the system
(\ref{p1q1sys}) is easily integrated and we arrive at the following
expressions for the generating functions $f_1$, $g_1$,
\begin{eqnarray}\label{f1g1}
&&f_1(z)=p_1(z)f_0(z),\quad
f_0(z)=u_0e^{i\frac{(\nu')^2}{8}z},\nonumber
\\
&&p_1(z)=\bigl(-\frac{i\nu\nu''}{4}-\frac{\nu''}{8}-
\frac{\nu'u_0'}{4u_0}\bigr)z-i\frac{(\nu')^2\nu''}{32}z^2,\nonumber
\\
&&g_1(z)=q_1(z)g_0(z),\quad
g_0(z)=v_0e^{-i\frac{(\nu')^2}{8}z},\nonumber
\\
&&q_1(z)=
\bigl(\frac{i\nu\nu''}{4}-\frac{\nu''}{8}-
\frac{\nu'v_0'}{4v_0}\bigr)z+i\frac{(\nu')^2\nu''}{32}z^2.
\end{eqnarray}
Finally, expanding the generating functions in series of $z$, we find the
coefficients $u_{n,2n-1}$ and $v_{n,2n-1}$,
\begin{eqnarray}\label{un2n-1}
&&\hskip-10pt
u_{1,1}=u_0\bigl(-\frac{i\nu\nu''}{4}-\frac{\nu''}{8}-
\frac{\nu'u_0'}{4u_0}\bigr),\quad
v_{1,1}=v_0\bigl(\frac{i\nu\nu''}{4}-\frac{\nu''}{8}-
\frac{\nu'v_0'}{4v_0}\bigr),
\\
&&\hskip-10pt
u_{n,2n-1}=-2u_0
\frac{i^{n-1}(\nu')^{2(n-1)}}{8^n(n-2)!}\Bigl(\nu''+
\frac{1}{n-1}\bigl(\frac{\nu''}{2}+i\nu\nu''+\frac{\nu'u_0'}{u_0}\bigr)
\Bigr),\quad
n\geq2,\nonumber
\\
&&\hskip-10pt
v_{n,2n-1}=-2v_0
\frac{(-i)^{n-1}(\nu')^{2(n-1)}}{8^n(n-2)!}\Bigl(
\nu''+\frac{1}{n-1}\bigl(\frac{\nu''}{2}-i\nu\nu''+
\frac{\nu'v_0'}{v_0}\bigr)\Bigr),\quad
n\geq2.\nonumber
\end{eqnarray}

Generating functions $f_j(z)$, $g_j(z)$ for $u_{n,2n-j}$, $v_{n,2n-j}$,
$j\geq2$, satisfy the differential system (\ref{fjgjsys}). Similarly to the
case $j=1$ above, let us introduce the auxiliary functions $p_j$ and $q_j$,
\begin{equation}\label{pjqjdef}
p_j=\frac{f_j}{f_0},\quad
q_j=\frac{g_j}{g_0}.
\end{equation}
In the terms of these functions, the system (\ref{fjgjsys}) reads,
\begin{equation}\label{pjqjsys}
\partial_zp_j=\frac{i\nu}{z}(p_j+q_j)+a_j,\quad
\partial_zq_j=-\frac{i\nu}{z}(p_j+q_j)+b_j,
\end{equation}
where
\begin{eqnarray}\label{ajbj}
&&a_j=2\partial_zp_{j-1}+
\bigl(\frac{i(\nu')^2}{4}-\frac{\nu''}{8}-
\frac{j-1}{z}\bigr)p_{j-1}-\nonumber
\\
&&-\frac{\nu'}{4}\frac{(p_{j-1}f_0)'}{f_0}-
i\frac{(p_{j-2}f_0)''}{8f_0}+
\frac{i\nu}{z}\sum_{k,l,m=0\atop k+l+m=j}^{j-1}p_kp_lq_m,\nonumber
\\
&&b_j=2\partial_zq_{j-1}+
\bigl(-\frac{i(\nu')^2}{4}-\frac{\nu''}{8}-
\frac{j-1}{z}\bigr)q_{j-1}-\nonumber
\\
&&-\frac{\nu'}{4}\frac{(q_{j-1}g_0)'}{g_0}+
i\frac{(q_{j-2}g_0)''}{8g_0}-
\frac{i\nu}{z}\sum_{k,l,m=0\atop k+l+m=j}^{j-1}p_kq_lq_m.
\end{eqnarray}
With the initial condition $p_j(0)=q_j(0)=0$, the system is easily
integrated and uniquely determines the functions $p_j(z)$, $q_j(z)$,
\begin{eqnarray}\label{pjqj}
&&p_j(z)=\int_0^za_j(\zeta)\,d\zeta+
i\nu\int_0^z\frac{d\zeta}{\zeta}
\int_0^{\zeta}d\xi(a_j(\xi)+b_j(\xi)),\nonumber
\\
&&q_j(z)=\int_0^zb_j(\zeta)\,d\zeta-
i\nu\int_0^z\frac{d\zeta}{\zeta}
\int_0^{\zeta}d\xi(a_j(\xi)+b_j(\xi)).
\end{eqnarray}
These equations with expressions (\ref{ajbj}) together establish the
recursion relation for the functions $p_j(z)$, $q_j(z)$. In terms of 
$p_j(z)$ and $q_j(z)$, expansion (\ref{as_expansion}) reads
\begin{eqnarray}\label{as_expansion_pjqj}
&&g_+=e^{i\frac{x^2}{2t}-(\frac{1}{2}+i\nu)\ln4t+
i\frac{(\nu')^2\ln^24t}{8t}}
u_0\sum_{j=0}^{\infty}
\frac{p_j\bigl(\frac{\ln^24t}{t}\bigr)}{\ln^j4t},\nonumber
\\
&&g_-=e^{-i\frac{x^2}{2t}-(\frac{1}{2}-i\nu)\ln4t-
i\frac{(\nu')^2\ln^24t}{8t}}
v_0\sum_{j=0}^{\infty}
\frac{q_j\bigl(\frac{\ln^24t}{t}\bigr)}{\ln^j4t}.
\end{eqnarray}

Let $a_j(z)$ and $b_j(z)$ be polynomials of degree $M$ with the zero $z=0$ of
multiplicity $m$,
$$
a_j(z)=\sum_{k=m}^Ma_{jk}z^k,\quad
b_j(z)=\sum_{k=m}^Mb_{jk}z^k.$$
Then the functions $p_j(z)$ and $q_j(z)$ (\ref{pjqj}) are polynomials of
degree $M+1$ with a zero at $z=0$ of multiplicity $m+1$,
\begin{eqnarray}\label{pjqj_sum}
&&p_j(z)=\sum_{k=m+1}^{M+1}\frac{1}{k}\bigl(
a_{j,k-1}+\frac{i\nu}{k}(a_{j,k-1}+b_{j,k-1})\bigr)z^k,\nonumber
\\
&&q_j(z)=\sum_{k=m+1}^{M+1}\frac{1}{k}\bigl(
b_{j,k-1}-\frac{i\nu}{k}(a_{j,k-1}+b_{j,k-1})\bigr)z^k.
\end{eqnarray}
On the other hand, $a_j(z)$ and $b_j(z)$ are described in (\ref{ajbj}) as
the actions of the differential operators applied to the functions
$p_{j'}$, $q_{j'}$ with $j'<j$. Because $p_0(z)=q_0(z)\equiv1$ and $p_1(z)$, 
$q_1(z)$ are polynomials of the second degree and a single zero at $z=0$, 
cf.\ (\ref{f1g1}), it easy to check that $a_2(z)$ and $b_2(z)$ are 
non-homogeneous polynomials of the third degree such that
\begin{eqnarray}\label{a2b2coeff}
&&a_{2,3}=-\frac{(\nu')^4(\nu'')^2}{2^{10}}(2-i\nu),\quad
b_{2,3}=-\frac{(\nu')^4(\nu'')^2}{2^{10}}(2+i\nu),
\\
&&a_{2,0}=-\frac{i\nu\nu''}{4}-\frac{\nu''}{8}-\frac{\nu'u_0'}{4u_0}-
\frac{iu_0''}{8u_0},\quad
b_{2,0}=\frac{i\nu\nu''}{4}-\frac{\nu''}{8}-\frac{\nu'v_0'}{4v_0}+
\frac{iv_0''}{8v_0}.\nonumber
\end{eqnarray}
Thus $p_2(z)$ and $q_2(z)$ are polynomials of the fourth
degree with a single zero at $z=0$. Some of their coefficients are
\begin{eqnarray}\label{p2q2coeff}
&&p_{2,4}=q_{2,4}=-\frac{(\nu')^4(\nu'')^2}{2^{11}},
\\
&&p_{2,1}=-\frac{i(\nu')^2}{4}-
(1+2i\nu)\bigl(\frac{\nu''}{4}+
\frac{iu_0''}{8u_0}\bigr)-\frac{\nu(u_0')^2}{4u_0^2},\nonumber
\\
&&q_{2,1}=\frac{i(\nu')^2}{4}-
(1-2i\nu)\bigl(\frac{\nu''}{4}-
\frac{iv_0''}{8v_0}\bigr)-\frac{\nu(v_0')^2}{4v_0^2}.\nonumber
\end{eqnarray}

\begin{lemma}\label{lemma4}
Functions $p_j(z)$, $q_j(z)$, $j\in{\Bbb Z}_+$, defined in (\ref{pjqjdef})
are polynomials of degree not exceeding $2j$. The zero of $p_j(z)$ and
$q_j(z)$ at $z=0$ has multiplicity no less that
$m_j=\bigl[\frac{j+1}{2}\bigr]$.
\end{lemma}
\proof
The assertion holds true for $j=0,1,2$. Let it be correct for
$\forall j<j'$. Then $a_{j'}(z)$ and $b_{j'}(z)$ are defined as the sum of
polynomials. The maximal degrees of such polynomials are
$\deg\bigl((p_{j'-1}f_0)'/f_0\bigr)=2j'-1$,
$\deg\bigl((q_{j'-1}g_0)'/g_0\bigr)=2j'-1$, and
$$
\deg\Bigl(\frac{1}{z}
\sum_{\alpha,\beta,\gamma=0\atop\alpha+\beta+\gamma=j'}^{j'-1}
p_{\alpha}p_{\beta}q_{\gamma}\Bigr)=2j'-1,\quad
\deg\Bigl(\frac{1}{z}
\sum_{\alpha,\beta,\gamma=0\atop\alpha+\beta+\gamma=j'}^{j'-1}
p_{\alpha}q_{\beta}q_{\gamma}\Bigr)=2j'-1.$$
Thus $\deg a_{j'}(z)=\deg b_{j'}(z)\leq2j'-1$, and
$\deg p_{j'}(z)=\deg q_{j'}(z)\leq2j'$.

Multiplicity of the zero at $z=0$ of $a_{j'}(z)$ and $b_{j'}(z)$ is no less
than the minimal multiplicity of the summed polynomials in (\ref{ajbj}), but
the minor coefficients of the polynomials $2\partial_zp_{j'-1}$ and
$-(j-1)p_{j'-1}/z$, as well as of $2\partial_zq_{j'-1}$ and
$-(j-1)q_{j'-1}/z$ may cancel each other. Let $j'=2k$ be even. Then
$$
m_{j'}=\min\bigl\{m_{j'-1};\,m_{j'-2}+1;\,
\min_{\alpha,\beta,\gamma=0,\dots,j'-1\atop
\alpha+\beta+\gamma=j'}\bigl(m_{\alpha}+m_{\beta}+m_{\gamma}\bigr)\bigr\}=
\frac{j'}{2}=\bigl[\frac{j'+1}{2}\bigr].$$
Let $j'=2k-1$ be odd. Then $2m_{j'-1}-(j'-1)=0$, and
$$
m_{j'}=\min\bigl\{m_{j'-1}+1;\,m_{j'-2}+1;\,
\min_{\alpha,\beta,\gamma=0,\dots,j'-1\atop
\alpha+\beta+\gamma=j'}\bigl(m_{\alpha}+m_{\beta}+m_{\gamma}\bigr)\bigr\}=
\frac{j'+1}{2}.$$
\endproof

We arrive at the following expansions of the polynomials $p_j(z)$,
$q_j(z)$,
\begin{equation}\label{pjqjexpansion}
p_j(z)=\sum_{k=[\frac{j+1}{2}]}^{2j}p_{j,k}z^k,\quad
q_j(z)=\sum_{k=[\frac{j+1}{2}]}^{2j}q_{j,k}z^k.
\end{equation}
The use of (\ref{pjqjexpansion}) in (\ref{as_expansion_pjqj}) yields 
(\ref{as_modified}). On the other hand, using (\ref{pjqjexpansion}) in 
(\ref{pjqjdef}), we obtain the expressions for the generating functions 
$f_j(z)$, $g_j(z)$,
\begin{eqnarray}\label{fjgjexpansion}
&&f_j(z)=u_0\sum_{n=[\frac{j+1}{2}]}^{\infty}z^n
\sum_{k=\max\{0;n-2j\}}^{n-[\frac{j+1}{2}]}
p_{j,n-k}\frac{i^k(\nu')^{2k}}{8^kk!},\nonumber
\\
&&g_j(z)=v_0\sum_{n=[\frac{j+1}{2}]}^{\infty}z^n
\sum_{k=\max\{0;n-2j\}}^{n-[\frac{j+1}{2}]}
q_{j,n-k}\frac{(-i)^k(\nu')^{2k}}{8^kk!},
\end{eqnarray}
and the coefficients of our interest,
\begin{eqnarray}\label{uvn2n-j}
&&u_{n,2n-j}=u_0\sum_{k=\max\{0;n-2j\}}^{n-[\frac{j+1}{2}]}
p_{j,n-k}\frac{i^k(\nu')^{2k}}{8^kk!},\nonumber
\\
&&v_{n,2n-j}=v_0\sum_{k=\max\{0;n-2j\}}^{n-[\frac{j+1}{2}]}
q_{j,n-k}\frac{(-i)^k(\nu')^{2k}}{8^kk!}.
\end{eqnarray}
In particular, the leading asymptotic term of these coefficients as
$n\to\infty$ and $j$ fixed is given by
\begin{eqnarray}\label{uvn2n-j_asymp_jconst}
&&u_{n,2n-j}=u_0p_{j,2j}\frac{i^{n-2j}(\nu')^{2(n-2j)}}{8^{n-2j}(n-2j)!}
\bigl(1+{\cal O}(\frac{1}{n})\bigr),\nonumber
\\
&&v_{n,2n-j}=v_0q_{j,2j}\frac{(-i)^{n-2j}(\nu')^{2(n-2j)}}{8^{n-2j}(n-2j)!}
\bigl(1+{\cal O}(\frac{1}{n})\bigr).
\end{eqnarray}
Thus we have reduced the problem of the evaluation of the asymptotics of
the coefficients $u_{n,2n-j}$ $v_{n,2n-j}$ for large $n$ to the computation
of the leading coefficients of the polynomials $p_j(z)$, $q_j(z)$. In fact,
using (\ref{pjqj}) or (\ref{pjqj_sum}) and (\ref{ajbj}), it can be shown
that the coefficients $p_{j,2j}$, $q_{j,2j}$ satisfy the recurrence
relations
\begin{eqnarray}\label{ljmj_recurrence}
&&p_{j,2j}=-i\frac{(\nu')^2\nu''}{2^5 j}p_{j-1,2(j-1)}+
\frac{i\nu}{2j}\sum_{k,l,m=0\atop k+l+m=j}^{j-1}p_{k,2k}p_{l,2l}q_{m,2m}+
\nonumber
\\
&&\hskip-4pt
+\frac{\nu(\nu')^2\nu''}{2^6 j^2}(p_{j-1,2(j-1)}-q_{j-1,2(j-1)})-
\frac{\nu^2}{4j^2}\sum_{k,l,m=0\atop k+l+m=j}^{j-1}
p_{k,2k}(p_{l,2l}-q_{l,2l})q_{m,2m},\nonumber
\\
&&q_{j,2j}=i\frac{(\nu')^2\nu''}{2^5 j}q_{j-1,2(j-1)}-
\frac{i\nu}{2j}\sum_{k,l,m=0\atop k+l+m=j}^{j-1}p_{k,2k}q_{l,2l}q_{m,2m}-
\\
&&\hskip-4pt
-\frac{\nu(\nu')^2\nu''}{2^6 j^2}(p_{j-1,2(j-1)}-q_{j-1,2(j-1)})+
\frac{\nu^2}{4j^2}\sum_{k,l,m=0\atop k+l+m=j}^{j-1}
p_{k,2k}(p_{l,2l}-q_{l,2l})q_{m,2m}.\nonumber
\end{eqnarray}

Similarly, the coefficients $u_{n,0}$, $v_{n,0}$ for the non-logarithmic
terms appears from (\ref{uvn2n-j}) for $j=2n$, and are given simply by
\begin{equation}\label{uvn0asym}
u_{n,0}=u_0p_{2n,n},\quad
v_{n,0}=v_0q_{2n,n}.
\end{equation}
Thus the problem of evaluation of the asymptotics of the coefficients
$u_{n,0}$, $v_{n,0}$ for $n$ large is equivalent to computation of the
asymptotics of the minor coefficients in the polynomials $p_j(z)$, $q_j(z)$.
However, the last problem does not allow a straightforward solution because, 
according to (\ref{FG}), the sectional generating functions for the
coefficients $u_{n,0}$, $v_{n,0}$ are given by
$$
F(z,0)=\sum_{n=0}^{\infty}u_{n,0}z^n,\quad
G(z,0)=\sum_{n=0}^{\infty}v_{n,0}z^n,$$
and solve the separated Nonlinear Schr\"odinger equation
\begin{eqnarray}\label{FG_NLS0}
&&-iF_{\zeta}+izF_z=\nu F+\frac{1}{8}zF''+F^2G,\nonumber
\\
&&iG_{\zeta}-izG_z=\nu G+\frac{1}{8}zG''+FG^2.
\end{eqnarray}

\section{Discussion}

Our consideration based on the use of generating functions of different
types reveals the asymptotic behavior of the coefficients $u_{n,2n-j}$,
$v_{n,2n-j}$ as $n\to\infty$ and $j$ fixed for the long time asymptotic
expansion (\ref{as_expansion}) of the generic solution of the sNLS equation
(\ref{NLS}). The leading order dependence of these coefficients on $n$ is 
described by the ratio $\frac{a^n}{(n-2j)!}$.

However, the behavior of the coefficients $u_{n,k}$, $v_{n,k}$ corresponding
to other directions in $(n,k)$-plane can be dramatically different. For
instance, the generating functions for $u_{n,0}$, $v_{n,0}$ solving equation
(\ref{FG_NLS0}) are related to the so-called isomonodromy solutions \cite{I}
of the system (\ref{NLS}). These solutions are characterized by the formal
asymptotic expansion (\ref{as_expansion}) without log-terms. Actually, it
takes place when parameter $\nu=-u_0v_0$ in (\ref{as_expansion}) is
independent of $\lambda_0=-x/2t$. Concerning the asymptotic behavior of
the coefficients $u_{n,0}$, $v_{n,0}$ as $n\to\infty$, it is the plausible
conjecture that $u_{n,0},v_{n,0}\sim c^n\Gamma(\frac{n}{2}+d)$. The
investigation of the Riemann-Hilbert problem for the sNLS equation yielding
this estimate will be published elsewhere.

\bigskip
{\bf Acknowledgments.} We are grateful to the support of NSF Grant
PHY-9988566. We also express our gratitude to P.~Deift, A.~Its and X.~Zhou
for discussions. A.~K.\ was partially supported by the Russian Foundation
for Basic Research under grant 99-01-00687. He is also grateful to the staff
of C.~N.~Yang Institute for Theoretical Physics of the State University of
New York at Stony Brook for hospitality during his visit when this work was
done.

\ifx\undefined\bysame
\newcommand{\bysame}{\leavevmode\hbox to3em{\hrulefill}\,}
\fi

\end{document}